\documentclass[twocolumn,prb,superscriptaddress]{revtex4}
\usepackage{graphicx}
\usepackage{epsfig}

\def\beq{\begin{equation}}
\def\eeq{\end{equation}}
\begin{document}



\title{ Unusual X-ray excited luminescence spectra of NiO suggestive of a self-trapping of the \emph{d}-\emph{d} charge transfer exciton }

\author{V.I.~Sokolov}
\affiliation{Institute of Metal Physics UD RAS, S. Kovalevskaya Str. 18, 620990, Ekaterinburg, Russia}
\author{V.A.~Pustovarov}
\affiliation{Ural Federal University, Mira Str., 19, 620002 Ekaterinburg, Russia}
\author{V.N.~Churmanov}
\affiliation{Ural Federal University, Mira Str., 19, 620002 Ekaterinburg, Russia}
\author{V.Yu.~Ivanov}
\affiliation{Ural Federal University, Mira Str., 19, 620002 Ekaterinburg, Russia}
\author{N.B.~Gruzdev}
\affiliation{Institute of Metal Physics UD RAS, S. Kovalevskaya Str. 18, 620990, Ekaterinburg, Russia}
\author{P.S.~Sokolov}
\affiliation{Lomonosov Moscow State University, 119991, Moscow, Russia}
\author{A.N.~Baranov}
\affiliation{Lomonosov Moscow State University, 119991, Moscow, Russia}
\author{A.S.~Moskvin}
\affiliation{Department of Theoretical Physics, Institute of Natural Science, Ural Federal University, Lenin Str., 51, 620083 Ekaterinburg, Russia}
\date{\today}

\begin{abstract}
Luminescence spectra of  NiO  have been investigated under vacuum ultraviolet (VUV) and soft X-ray (XUV) excitation. Photoluminescence (PL) spectra show broad emission bands centered at about 2.3 and 3.2 eV. The PL excitation (PLE) spectral evolution and lifetime measurements  reveal that two mechanisms with short and long  decay times, attributed to the d($e_g$)-d($e_g$) and p($\pi$)-d  charge transfer (CT) transitions in the range 4-6\,eV, respectively, are responsible for the observed emissions, while the most intensive p($\sigma$)-d CT transition at 7\,eV appears to be a weak if any PL excitation mechanism. The PLE spectra recorded in the 4-7\,eV range agree with the RIXS and reflectance data. 
Making use of the XUV excitation allows us to avoid the predominant role of the surface effects in luminescence and reveal bulk luminescence with puzzling well isolated doublet of very narrow lines with close energies near 3.3\,eV  characteristic for recombination transitions in self-trapped \emph{d}-\emph{d} CT excitons formed by coupled Jahn-Teller Ni$^+$ and Ni$^{3+}$ centers. This conclusion is supported both by a comparative analysis of the luminescence spectra for NiO and solid solutions Ni$_{x}$Zn$_{1-x}$O, and by a comprehensive cluster model assignement of different \emph{p}-\emph{d} and \emph{d}-\emph{d} CT transitions,  their relaxation channels. To the best of our knowledge it is the first observation of the self-trapping for \emph{d}-\emph{d} CT excitons.
Our paper shows the time resolved luminescence measurements provide an instructive tool for elucidation of the \emph{p}-\emph{d} and \emph{d}-\emph{d} CT excitations and their relaxation in 3d oxides.  

\end{abstract}

\maketitle

\section{Introduction}
Explaining the electronic properties of transition metal monoxides is one of the long-standing problems in the condensed matter physics. Nickel monoxide NiO with its rather simple rocksalt structure, a large insulating gap and an antiferromagnetic  ordering temperature of T$_N$\,=\,523\,K has been attracting many physicists as a prototype oxide for this problem. This strongly correlated electron material has played and is playing a very important role in clarifying the electronic structure and understanding the rich physical properties of 3d compounds. Recent discoveries of a giant low frequency dielectric constant, bistable resistance switching,  and other unconventional properties\,\cite{NiO} have
led to a major resurgence of interest in NiO.

However, despite several decades of studies\,\cite{Brandow} there is still no literature consensus on the detailed electronic structure of NiO. 
Different energy gap ($E_g$) values between 3.7 and 5\,eV\,\cite{photoconductivity,Hiraoka} are reported for NiO and reliable $E_g$ still remains to be settled.
Conventional band theories which stress the delocalized nature of electrons cannot
explain this large gap and predict NiO to be metallic. NiO has long been viewed as a prototype "Mott insulator",\,\cite{Brandow} with gap formed by intersite \emph{d}-\emph{d} charge transfer (CT) transitions, however, this view was later replaced by that of a "CT insulator"\,\cite{ZSA} with gap formed by  \emph{p}-\emph{d} CT transitions.
Most recent RIXS and optical reflectivity measurements\,\cite{Hiraoka} showed that the CT band peaked near 4-5\,eV reveales a discernible q-dispersion of its energy typical for Mott-Hubbard \emph{d}-\emph{d} CT transitions while intensive CT band at higher energy peaked near 7-8\,eV   reveales only intensity dispersion without any visible dispersion of the energy that is typical for intra-center \emph{p}-\emph{d} CT transitions. Nevertheless, to date we have no comprehensive assignment of different spectral features in NiO to \emph{p}-\emph{d} or \emph{d}-\emph{d} CT transitions.


Although the optical absorption of NiO have been studied experimentally with some detail, their optical emission properties have been scarcely investigated. However, luminescence spectroscopy can give a profound insight into the electronic structure, electron-hole excitations and their relaxation in the lattice. Different measurements performed with UV excitation below and near optical gap\,\cite{Diaz-Guerra,Volkov,Mochizuki} point to a broad luminescence band in the region 2-3.5\,eV. The temperature behavior indicates that the broad PL spectra of NiO consist of several components with relative intensity dependent on the temperature and excitation energy.
 The low-energy excitation ($E_{exc}\approx 2.4$\,eV) below the charge-transfer gap stimulates a photoemission  in single-crystal NiO with two maxima, one at 1.5-1.6\,eV and a larger one at 2.2-2.3\,eV\,\cite{Mironova}.
Green luminescence band with a maximum around 2.25\,eV has been observed in nanoclustered  NiO  at $E_{exc}\leq 2.95$\,eV\,\cite{Volkov}. The observed emission bands in the visible and near infrared spectral ranges are usually attributed to Ni$^{2+}$ intrasite, or crystal field \emph{d}-\emph{d} transitions. In particular, the main green luminescence band peaked near 2.3\,eV is attributed to a Stokes-shifted  ${}^1T_{2g}(D)\rightarrow {}^3A_{2g}(F)$ transition while the low-energy band peaked near 1.5\,eV is related to ${}^1E_{g}(D)\rightarrow {}^3A_{2g}(F)$ transition\,\cite{Mironova} (see Fig.\,1 for the spectrum of the crystal field \emph{d}-\emph{d} transitions\,\cite{Chrenko}).
\begin{figure}[t]
\includegraphics[width=8.5cm,angle=0]{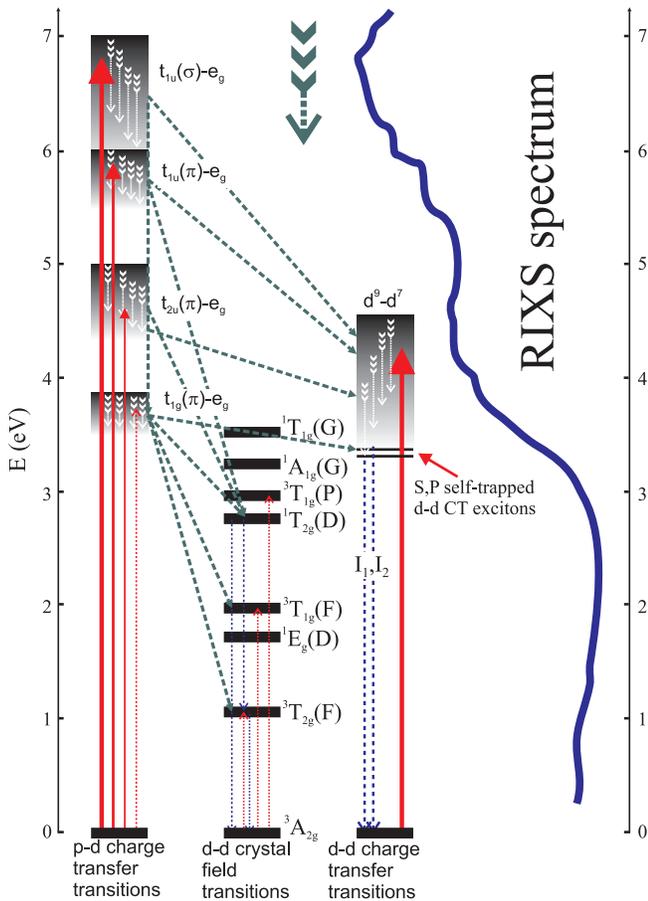}
\caption{(Color online) Spectra of the \emph{d}-\emph{d}, \emph{p}-\emph{d} CT transitions and intracenter crystal field \emph{d}-\emph{d} transitions in NiO. Strong dipole-allowed  $\sigma -\sigma$ \emph{d}-\emph{d} and \emph{p}-\emph{d} transitions are shown  by thick solid
arrows;  weak dipole-allowed  $\pi -\sigma$ \emph{p}-\emph{d} transitions by thin solid
arrows; weak dipole-forbidden low-energy transitions by thin dashed arrows,
respectively. Dashed lines point to different EH relaxation channels, dotted lines point to PL transitions. Spectrum of the crystal field transitions is reproduced from Ref.\,\onlinecite{Chrenko}. Right hand side reproduces a fragment of the RIXS spectra for NiO\,\cite{Duda}.} \label{fig1}
\end{figure}
However, the photoluminescence spectra of bulk single crystals and ceramics of NiO under 3.81\,eV photoexcitation\,\cite{Mochizuki} suggestive the band gap excitation  have revealed in addition to the green band a more intensive broad violet PL band with a maximum around 3\,eV. The band was related to a \emph{p}-\emph{d} charge transfer. 
Radiative recombination of carriers in powdered pellets of NiO under UV excitation  with $E_{exc}=4.43$\,eV (280\,nm) higher than the CT gap consists at 10\,K of a broad intensive band peaked at 2.8\,eV with a shoulder centered at about 3.2\,eV\,\cite{Diaz-Guerra}. 
 It is clearly appreciated that the 2.8\,eV emission band decreases with increasing temperature, while the 3.2\,eV band follows a strictly opposite trend and becomes the dominant emission at about 150\,K. Furthermore, the 3.2\,eV band reveals a two-peak structure clearly visible at elevated temperatures. 
 Different kinetic properties, seemingly different temperature behavior\,\cite{Diaz-Guerra} point to different relaxation channels governing the green and violet luminescence.

 The comparison of the PL spectra with the catodoluminescence (CL) spectra of the same samples\,\cite{Diaz-Guerra} reveals differences in the relative intensity of the excited emissions.
The 2.8 and 3.2\,eV PL bands appear in the CL spectra  as shoulders of a main broad green 2.4\,eV emission band. However, for NiO samples annealed in vacuum both the 2.4 and  3.2\,eV bands have a comparable spectral weight and clearly visible multipeak structure.   The green 2.4\,eV  band is visible in time-resolved PL spectra recorded at 10 K for different delay times\,\cite{Diaz-Guerra}. 
At variance with Ref.\,\onlinecite{Mochizuki} the PL and CL emission bands observed in this work\,\cite{Diaz-Guerra} were attributed to the crystal field \emph{d}-\emph{d} transitions.

To the best of our knowledge the photoexcitation of PL has been restricted by $E_{exc}=4.43$\,eV\,\cite{Diaz-Guerra} with no inspection of the PL photoexcitation over the CT band, though such a study can be an instructive tool to elucidate the mechanism of the CT transitions and the spectral selectivity of the PL.  
It is worth noting that all the studies of the PL in NiO point to a special role of different defects and the surface induced local non-cubic distortions in photoemission enhancement and a remarkable inhomogeneous broadening of the PL bands. Indeed, a most effective  absorption of photons with the energy $\hbar\omega \sim E_g$ occurs in a thin (10-20\,nm) surface layer with more or less distorted symmetry and enhanced defect concentration. In other words, the 
UV photoexcitation cannot stimulate the bulk luminescence mirroring the fundamental material properties.
These issues did motivate our studies of the photoluminescence spectra in NiO under high-energy excitation with making use of the both VUV and soft X-ray time-resolved PL excitation technique.

\section{Experimental results}
The PL measurements were made on the samples of NiO  and several solid solutions Ni$_{1-x}$Zn$_x$O ($x$\,=\,0.2, 0.3, and 0.6) with rock salt crystal structure.
As starting material we have used the commercially available powder of NiO (99\%; Prolabo) and ZnO (99.99\%; Alfa Aesar) which has been pressed into pellets under pressure of about 1250 bar and placed into gold capsules. Quenching experiments at 7.7\,GPa and 1000-1100\,K have been performed using a toroid-type high-pressure apparatus. Detailes of experimental technique and calibration are described elsewhere\,\cite{Baranov}. Electron microscopy analysis shows the samples to be dense poreless oxide ceramics with rock salt cubic structure and grain size of about 10-20\,$\mu m$. The NiO and  Ni$_{0.3}$Zn$_{0.7}$O ceramic samples has been stired and pressed into cellulose to enhance the luminescence intensity.
\begin{figure}[t]
\includegraphics[width=8.5cm,angle=0]{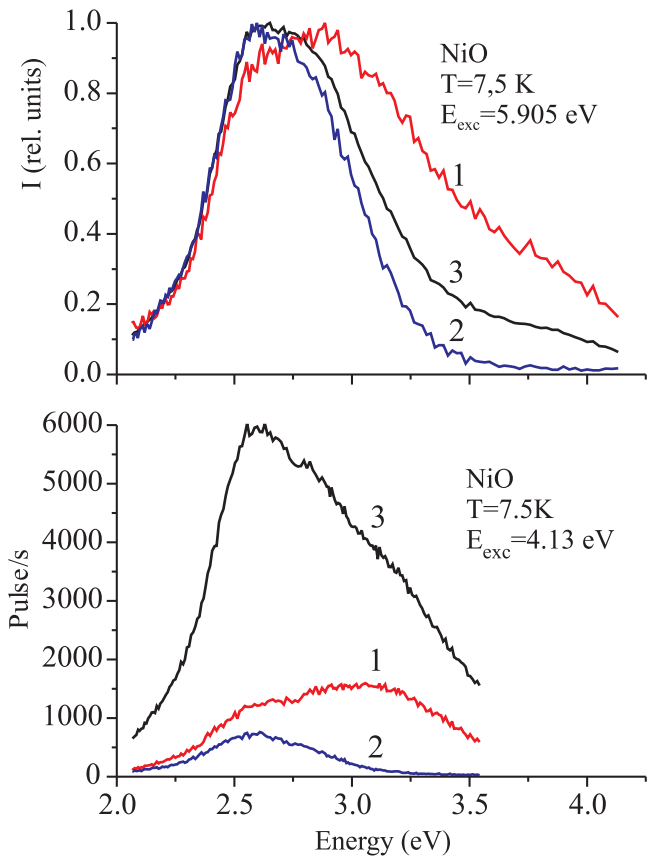}
\caption{(Color online) Time-resolved PL spectra  (VUV - excitation) of  NiO: 1 - fast window; 2 - slow window; 3 - time-integrated spectrum.} \label{fig2}
\end{figure}
The PL spectra have been excited by the synchrotron radiation (SR) with 1\,ns pulses running with 96\,ns intervals. The PL and PLE spectra were recorded in 2-3.5\, and 4-7\,eV range, respectively\,\cite{DESY}.  
The measurements of PL spectra under soft X-ray (XUV) excitation were made on a SUPERLUMI station (HASYLAB (DESY), Hamburg) using an ARC Spectra Pro-308i monochromator and  R6358P Hamamatsu photomultiplier. The time-resolved PL and PLE spectra, the PL decay kinetics in 1-80\,ns interval  were measured in two time windows:  fast with a delay time $\delta t_1 = 0.6$\,ns, span of windows $\Delta t_1 = 2.3$\,ns; and slow delay time $\delta t_2 = 58$\,ns, $\Delta t_2 = 14$\,ns. The time-resolved PL spectra  as well as the PL decay kinetics under XUV excitation has been measured on a BW3 beamline by a VUV monochromator (Seya-Namioka scheme) equipped with microchannel plate-photomultiplier (MCP 1645, Hamamatsu). The parameters of time windows: $\delta t$\,=0.1\,ns, $\Delta t$\,=5.7\,ns. The temporal resolution of the whole detection system was 250\,ps. The temporary interval between SR excitation pulses is equal 96\,ns. 

\begin{figure}[t]
\includegraphics[width=8.5cm,angle=0]{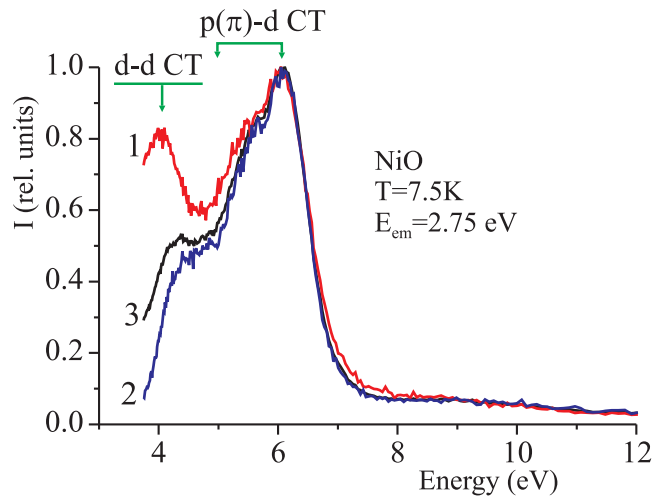}
\caption{(Color online) Time-resolved PLE spectra  of  NiO: 1 - fast window; 2 - slow window; 3 - time-integrated spectrum. Arrows point to $d(e_g$)-$d(e_g$) and $p(\pi$)-$d$ ($t_{2u}(\pi )\rightarrow e_g$ and $t_{1u}(\pi )\rightarrow e_g$) CT transitions which are most effective in the luminescence excitation.} \label{fig3}
\end{figure}
Photoluminescence spectra of NiO under VUV-excitation at two energies $\hbar\omega >E_g$ and different registration regimes are presented in Fig.\,2. The PL spectra have a complicated form depending on registration time and excitation energy.  Time-integrated spectrum (Fig.\,2, curve 3) has a maximum near 2.6\,eV and an extended high-energy tail which is more intensive for the 4.13\,eV  than for the 5.9\,eV  excitation.  Time-integrated spectrum at 4.13\,eV  excitation does  correlate with the PL spectrum of NiO  pressed powders observed in Ref.\,\onlinecite{Diaz-Guerra} at T=10\,K and $E_{exc}=4.43$\,eV. A significant difference in PL spectra is revealed for fast and slow windows. High-energy 5.9\,eV  excitation results in a sizeable enhancement of PL in the high-energy range 3-4\,eV for fast window, however, the  4.13\,eV  excitation gives rise to a more pronounced effect with the intensive maximum of the broad PL band  shifted to 3.15\,eV, while for slow window one observe a broad PL band peaked near 2.6\,eV. It is worth noting that the PL spectrum for fast window and 4.13\,eV  excitation correlates rather well with that in Ref.\,\onlinecite{Mochizuki} ($E_{exc}=3.81$\,eV, T=10\,K). 
\begin{figure}[t]
\includegraphics[width=8.5cm,angle=0]{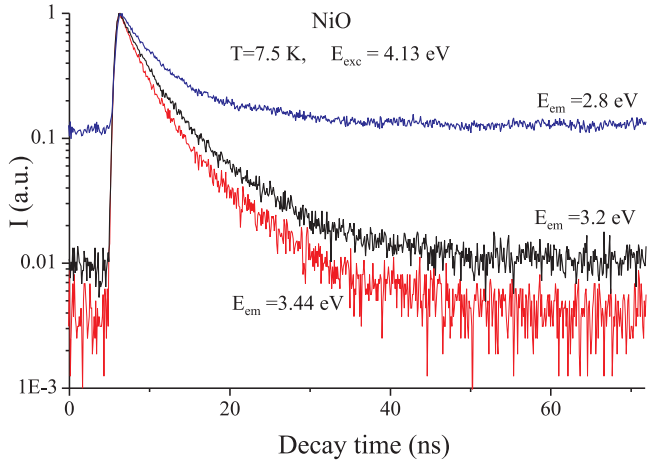}
\caption{(Color online) PL decay kinetics (VUV - excitation) of  NiO.} \label{fig4}
\end{figure}
The PL excitation spectrum shown in Fig.\,3 reveals a significant effect of different observation conditions. Time-integrated spectrum and the spectrum for slow window present a broad maximum near 6\,eV and a shoulder near 4.5\,eV.  For fast window the shoulder transforms into a narrow peak near 4\,eV comparable in magnitude with strong maximum near 6\,eV. 

It should be emphasized that the CT transition peaked near 4\,eV working only in the fast window is the most effective in excitation of the high-energy (violet) part of the PL spectrum. Interestingly, the PLE spectrum correlates both with the reflectance\,\cite{Powell},  electroreflectance\,\cite{electroreflectance}, and  RIXS\,\cite{Duda} spectra of NiO in the region 4-6\,eV.
The PL decay kinetics is shown in Fig.\,4 for  $E_{exc}$\,=\,4.13\,eV and different emission energies. The decay curves were approximated by a sum of two exponentials and a pedestal which describes a slow microsecond decay: $I(t)=y_0 + y_1\,exp (-t/\tau_1) + y_2\,exp (-t/\tau_2) $, where
$\tau_1$\,=\,3.4\,ns, $\tau_2$\,=\,14.6\,ns, $y_0$\,=\,0.11 ($E_{exc}$\,=\,4.13\,eV, $E_{em}$\,=\,2.8\,eV); $\tau_1$\,=\,2.3\,ns, $\tau_2$\,=\,7.2\,ns, $y_0$\,=\,0.003 ($E_{exc}$\,=\,4.13\,eV, $E_{em}$\,=\,3.44\,eV); $\tau_1$\,=\,2.3\,ns, $\tau_2$\,=\,16.7\,ns, $y_0$\,=\,0.006 ($E_{exc}$\,=\,5.9\,eV, $E_{em}$\,=\,2.8\,eV).  Thus, time-resolved  PL and PLE spectra, as well as the PL decay kinetics point to two relaxation processes in NiO with a characteristic times of nanoseconds and ten of nanosecond, respectively. Part of energy is emitted in micro- or millisecond range.   
\begin{figure}[t]
\includegraphics[width=8.5cm,angle=0]{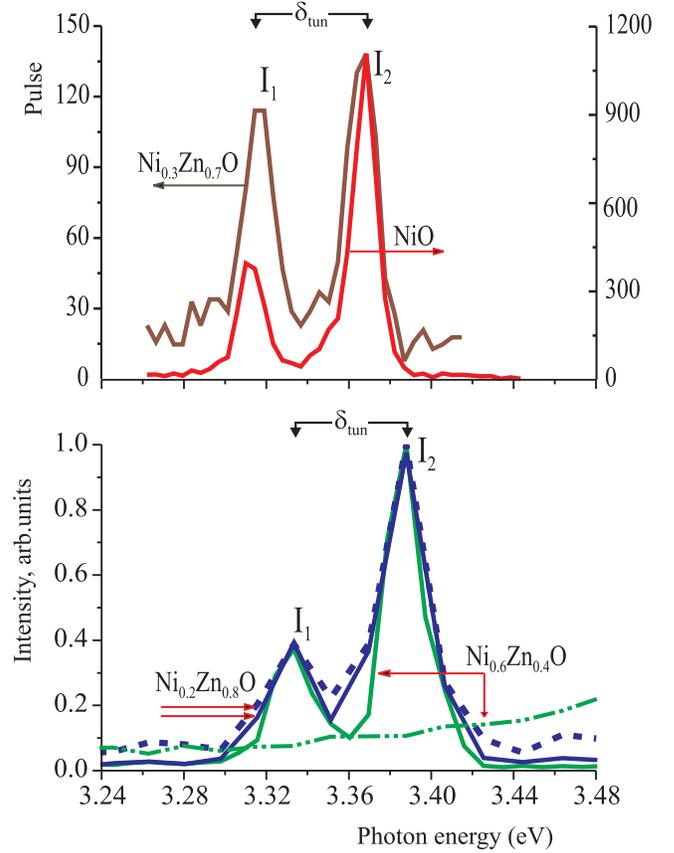}
\caption{(Color online) XUV excited luminescence spectra of NiO and solid solutions Ni$_x$Zn$_{1-x}$O (fast window). Upper panel: Luminescence spectra of the cellulose coated NiO and  Ni$_{0.3}$Zn$_{0.7}$O samples under XUV excitation with energy $E_{exc}$\,=\,130\,eV at T\,=7.2\,K. Bottom panel: Low-temperature (T\,=\,7.5\,K) luminescence spectra of the Ni$_{0.2}$Zn$_{0.8}$O and  Ni$_{0.6}$Zn$_{0.4}$O cellulose free samples  under XUV excitation with energy $E_{exc}$\,=\,130\,eV (solid line) and $E_{exc}$\,=\,450\,eV (dashed line). Luminescence spectra of Ni$_{0.6}$Zn$_{0.4}$O under XUV excitation with energy $E_{exc}$\,=\,130\,eV at T\,=\,7.2\,K (solid line) and room temperature (dash-and-dotted line).}   \label{fig5}
\end{figure}

Luminescence spectra of NiO under XUV excitation with energy $E_{exc}$\,=\,130\,eV and fast window opening by 100\,ps  after the excitation impulse start are presented in Fig.\,5. The XUV excited luminescence reveals puzzling spectral features with two close and very narrow lines $I_1$ and $I_2$ with a short decay-time $\tau <$\,400\,ps peaked for NiO sample at 3.310\,eV (linewidth 17\,meV) and 3.369\,eV (linewidth 13\,meV), respectively, mounted on a weak broad structureless pedestal in the 2.5-4\,eV range which is actually observed only for slow window. 
To the best of our knowledge, such an unusual luminescence has not been observed to date either in NiO or other 3d oxides. From the other hand, the well isolated $I_1$-$I_2$ doublet in the XUV excited luminescence seems to be a close relative of the broad high-energy (violet)  band in PL spectra peaked near 3.2\,eV. Dramatic difference in violet luminescence spectra under XUV and VUV excitation can be explained if to account for different penetration depth of VUV and XUV quanta. XUV  excitation stimulates the bulk luminescence mirroring the fundamental material properties while the UV photoexcitation stimulates thin surface layers which irregularities give rise to a strongly enhanced and inhomogeneously broadened luminescence.  
To examine the origin of the unconventional $I_1$-$I_2$ doublet and make more reasonable suggestions about its nature   we have made the measurements of the XUV excited luminescence for solid solution Ni$_{0.3}$Zn$_{0.7}$O. As in NiO we observed the $I_1$-$I_2$ doublet actually with the same energies and close linewidths. However, the integral intensity of the $I_1$-$I_2$ doublet in Ni$_{0.3}$Zn$_{0.7}$O is appeared to be almost ten times weaker than in NiO that points to  the relation of the $I_1$-$I_2$ doublet with an emission produced by somehow coupled pairs of Ni ions. To exclude conceivable parasitic effect of the cellulose coating we have made the measurements of the XUV excited luminescence for cellulose free ceramic samples of solid solutions Ni$_{0.2}$Zn$_{0.8}$O and  Ni$_{0.6}$Zn$_{0.4}$O (Fig.\,5, bottom panel). For the both samples we have observed the same $I_1$-$I_2$ doublet structure of the luminescence spectra with practically the same energy separation $\delta \approx $\,60\,meV and a small 20\,meV blue shift as compared with NiO and  Ni$_{0.3}$Zn$_{0.7}$O samples. Such a shift is believed to arise from small strains induced by coatings. Interestingly, the novel luminescence is clearly visible only at low temperatures: room temperature measurements do not reveal noticeable effect (see RT spectrum for Ni$_{0.6}$Zn$_{0.4}$O in Fig.\,5 typical for other samples). As it is seen in Fig.\,5 (bottom panel) the XUV excitation with higher energy $E_{exc}$\,=\,450\,eV does induce nearly the same $I_1$-$I_2$ doublet structure of the luminescence spectra.

\section{Discussion}
Charge carriers and excitons photogenerated in a crystal with strong electron-lattice interaction are known to relax to self-trapped states causing local lattice deformation and forming luminescence centers.
Despite the nature of radiative and nonradiative transitions in  strongly correlated 3\emph{d} oxides is far from full  understanding some reliable semiquantitative predictions can be made in frames of a simple cluster approach (see, e.g. Refs.\,\onlinecite{Moskvin-CT} and references therein).
The method provides a clear physical
picture of the complex electronic structure and the energy
spectrum, as well as the possibility of a quantitative modelling.
In a certain sense the cluster
calculations might provide a better description of the overall
electronic structure of  insulating  3d oxides  than the 
band structure calculations, mainly  due to a better account for correlation effects.
 Starting with octahedral NiO$_6$ complex with  the point symmetry group $O_h$ we deal with  five Ni\,3d and eighteen  oxygen O\,2p atomic
orbitals  
forming both hybrid Ni\,3d-O 2p  bonding and antibonding $e_g$ and $t_{2g}$
molecular orbitals (MO), and purely oxygen nonbonding $a_{1g}(\sigma)$, $t_{1g}(\pi)$,
$t_{1u}(\sigma)$, $t_{1u}(\pi)$, $t_{2u}(\pi)$ orbitals.
   Ground state of [NiO$_6$]$^{10-}$ cluster, or nominally Ni$^{2+}$ ion corresponds to $t_{2g}^6e_g^2$ configuration with the Hund ${}^3A_{2g}$ ground term.
   
   Typically for the octahedral MeO$_6$ clusters,\cite{Moskvin-CT} the non-bonding $t_{1g}(\pi)$ oxygen orbital has the
highest energy and forms the first electron removal oxygen state while other nonbonding oxygen $\pi$-orbitals, $t_{2u}(\pi)$, $t_{1u}(\pi)$, and $\sigma$-orbital $t_{1u}(\sigma)$ have lower energy with the energy separation $\sim$\,1\,eV in between (see Fig.\,1). 
   
 


The \emph{p}-\emph{d} CT transition in NiO$_{6}^{10-}$ center  is related with the transfer of O 2p electron to the partially filled
3d$e_g$-subshell with formation on the Ni-site of the $(t_{2g}^{6}e_{g}^{3})$ configuration of nominal Ni$^{+}$ ion isoelectronic to the well-known Jahn-Teller Cu$^{2+}$ ion. Yet actually instead of a single \emph{p}-\emph{d} CT transition we arrive at a series of O 2p$\gamma \rightarrow$ Ni 3d$e_g$ CT transitions forming a complex \emph{p}-\emph{d} CT band. It should be noted that each single electron transition gives rise to two many-electron transitions. The band starts with the dipole-forbidden $t_{1g}(\pi )\rightarrow e_g$, or ${}^{3}A_{2g}\rightarrow {}^{3}T_{1g},{}^{3}T_{2g}$ transitions, then includes two formally dipole-allowed so-called $\pi \rightarrow \sigma$ \emph{p}-\emph{d} CT transitions, weak $t_{2u}(\pi )\rightarrow e_g$, and  relatively strong $t_{1u}(\pi )\rightarrow e_g$ CT transitions, respectively, each giving rise to ${}^{3}A_{2g}\rightarrow {}^{3}T_{1u},{}^{3}T_{2u}$ transitions. Finally main \emph{p}-\emph{d} CT band is ended by 
the  strongest dipole-allowed $\sigma \rightarrow \sigma$   $t_{1u}(\sigma )\rightarrow e_g$ (${}^{3}A_{2g}\rightarrow {}^{3}T_{1u},{}^{3}T_{2u}$) CT transition. Above estimates predict the separation between partial \emph{p}-\emph{d} bands to be $\sim$\,1\,eV. Thus, if the most intensive CT band with a maximum around 7\,eV observed in RIXS spectra\,\cite{Duda,Hiraoka} to attribute to the  strongest dipole-allowed  O 2p$t_{1u}(\sigma )\rightarrow$ Ni 3d$e_g$ CT transition then one should expect the low-energy \emph{p}-\emph{d} CT counterparts with maximuma around 4, 5,  and 6\,eV\, respectively, which are related to dipole-forbidden $t_{1g}(\pi )\rightarrow e_g$, weak dipole-allowed  $t_{2u}(\pi )\rightarrow e_g$, and  relatively strong dipole-allowed  $t_{1u}(\pi )\rightarrow e_g$ CT transitions, respectively (see Fig.\,1). It is worth noting that  the $\pi \rightarrow \sigma$ \emph{p}-\emph{d} CT $t_{1u}(\pi)-e_{g}$
transition borrows a portion of intensity from the strongest dipole-allowed $\sigma \rightarrow \sigma$ $t_{1u}(\sigma )-e_g$ CT transition
because the $t_{1u}(\pi )$ and $t_{1u}(\sigma )$ states of the same symmetry are partly hybridized due to p-p covalency and overlap.
Interestingly that this assignement finds a strong support in the reflectance (4.9, 6.1, and 7.2\,eV for allowed \emph{p}-\emph{d} CT transitions )\,\cite{Powell} and, particularly, in electroreflectance spectra\,\cite{Powell,electroreflectance} which detect dipole-forbidden transitions. Indeed, the  spectra clearly point to a forbidden transition peaked near 3.7\,eV (missed in reflectance spectra) which thus  defines a \emph{p}-\emph{d} character of the optical CT gap and can be related with the on-set transition for the whole complex \emph{p}-\emph{d} CT band. It should be noted that a peak near 3.8\,eV has been observed in nonlinear absorption spectra of NiO\,\cite{Pisarev}.
 A rather strong p($\pi$)-d CT band peaked at 6.3\,eV is clearly visible in the absorption spectra of MgO:Ni\,\cite{Blazey}.
 
As a result of the \emph{p}-\emph{d} CT transition, a photo-generated electron localizes on a Ni$^{2+}$ ion forming Jahn-Teller 3d$^9$, or Ni$^{1+}$ configuration, while a photo-generated hole can move more or less itinerantly in the O 2p valence band determining the hole-like photoconductivity\,\cite{photoconductivity}. It is worth noting that any oxygen $\pi$-holes have larger effective mass than $\sigma$-holes, that results in the substantially distinct role of $p(\pi$)-$d$ and $p(\sigma$)-$d$ CT transitions both in photoconductivity\,\cite{remark} and the luminescence stimulation.   


The most effective channel of the recombinational relaxation for the spin-triplet \emph{p}-\emph{d} CT states $t_{2g}^6e_g^3\underline{\gamma}(\pi )$ ($\gamma (\pi)=t_{1g}(\pi),t_{2u}(\pi ),t_{1u}(\pi )$) implies the $\pi\rightarrow\pi$ transfer $t_{2g}\rightarrow\gamma(\pi )$ with formation of  excited spin-triplet ${}^{3}T_{1g}$ or ${}^{3}T_{2g}$ states of the $t_{2g}^5e_g^3$ configuration of Ni$^{2+}$ ion followed by final relaxation to lowest singlet terms ${}^{1}T_{2g}$ and ${}^{1}E_{g}$ producing green and red luminescence, respectively. Obviously, this relaxation is strongly enhanced by any symmetry breaking effects lifting or weakening the selection rules.
It means the three $\pi\rightarrow\sigma$ \emph{p}-\emph{d} CT transitions $t_{1g}(\pi)\rightarrow e_g$, $t_{2u}(\pi )\rightarrow e_g$, and $t_{1u}(\pi )\rightarrow e_g$ are expected to effectively stimulate the green luminescence due to large effective mass of $\pi$-oxygen holes and a well-localized and long-lived character of the \emph{p}-\emph{d} excitation. On the other hand,   the most intensive $\sigma\rightarrow\sigma$ \emph{p}-\emph{d} CT transition $t_{1g}(\sigma )\rightarrow e_g$ appears to be significantly less effective due to small effective mass of the $\sigma$-oxygen holes and relatively short lifetime of the respective unstable \emph{p}-\emph{d} CT state. It does explain the lack of the 7\,eV peak in the PLE excitation spectra (Fig.\,3).
 
Thus, the $p(\pi )$-$d$ CT transitions are believed to effectively stimulate the green luminescence of NiO, yet these cannot explain the origin of violet luminescence, in particular, specific $I_1$-$I_2$ doublet stimulated by XUV excitation which concentration behavior points to  participation of Ni pairs, or \emph{d}-\emph{d} CT transitions  rather than isolated NiO$_6$ centers.
Indeed, strong \emph{d}-\emph{d} CT, or Mott transitions provide an important contribution to the optical response of strongly correlated 3d oxides\,\cite{Moskvin-CT}.
In NiO one expects a strong  \emph{d}-\emph{d} CT transition related with the $\sigma -\sigma$-type  $e_g-e_g$ charge transfer $t_{2g}^6e_g^2+t_{2g}^6e_g^2\rightarrow t_{2g}^6e_g^3+t_{2g}^6e_g^1$ between $nnn$ Ni sites with the creation of electron  [NiO$_{6}$]$^{11-}$ and hole [NiO$_{6}$]$^{9-}$ centers (electron-hole dimer), or nominally Ni$^{+}$ and Ni$^{3+}$ ions. This unique anti-Jahn-Teller transition ${}^{3}A_{2g}+{}^{3}A_{2g}\rightarrow {}^{2}E_{g}+{}^{2}E_{g}$ creates a  \emph{d}-\emph{d} CT exciton self-trapped in the lattice due to electron-hole attraction and strong "double"\,Jahn-Teller effect for the both electron and hole centers. Exchange tunnel reaction  Ni$^{+}$+Ni$^{3+}$$\leftrightarrow$Ni$^{3+}$+Ni$^{+}$  due a two-electron transfer
gives rise to two symmetric (S- and P-) excitons (see Fig.\,6) having s- and p-symmetry, respectively, with energy separation $\delta =\delta_{tun}=2|t|$, where $t$ is the two-electron transfer integral which magnitude is of the order of Ni$^{2+}$-Ni$^{2+}$ exchange integral\,\cite{Moskvin-PRB-11}. Interestingly that  P-exciton is dipole-allowed while S-exciton is dipole-forbidden.
\begin{figure}[t]
\includegraphics[width=8.5cm,angle=0]{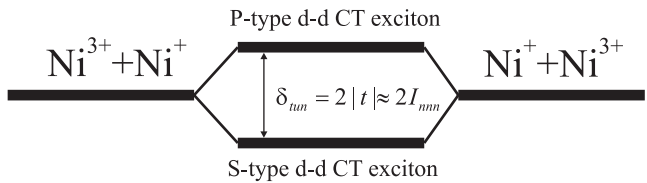}
\caption{Illustration to formation of S- and P-type \emph{d}-\emph{d} CT excitons in NiO.} \label{fig6}
\end{figure}

Strong dipole-allowed Franck-Condon $d(e_g)$-$d(e_g)$ CT transition in NiO manifests itself as a strong spectral feature near 4\,eV clearly visible by the RIXS spectra (4.5\,eV)\,\cite{Duda,Hiraoka}, the reflectance spectra (4.3\,eV)\,\cite{Powell}, the nonlinear absorption spectra (4.1-4.3\,eV)\,\cite{Pisarev}, and, particularly, in our PLE spectra (4.1\,eV, see Fig.\,3). Such a strong absorption near 4\,eV is beyond the predictions of the \emph{p}-\emph{d} CT model and indeed is lacking in absorption spectra of MgO:Ni\,\cite{Blazey}. 
 Accordingly, the CT gap in NiO is believed to be formed by a superposition of the electro-dipole forbidden \emph{p}-\emph{d} ($t_{1g}(\pi )\rightarrow e_g$) and allowed $d(e_g)$-$d(e_g)$ CT transitions with close energies.  
The energy of the S-P doublet of the self-trapped \emph{d}-\emph{d} CT exciton is expected to be $\sim$\,1.0-1.5\,eV lower than the energy of the $d(e_g)$-$d(e_g)$ peak, that is near 3\,eV, if to account for typical estimates for electron-hole attraction ($\sim$\,0.5\,eV) and Jahn-Teller stabilization energy ($\sim$\,0.5\,eV/one E-type state). It is worth noting that different optical reflectance and  absorption measurements\,\cite{Chrenko,Powell,Propach} did not reveal the $I_1$-$I_2$ doublet.

Obviously the recombination of the self-trapped \emph{d}-\emph{d} CT exciton, or, strictly speaking, of the S-P doublet  can explain all the features observed  in the XUV excited luminescence spectra (Fig.\,5). Weak line $I_1$ can be attributed to recombination of the dipole-forbidden S-exciton, while strong line $I_2$ with that for dipole-allowed P-exciton. The S-P separation ($\delta_{tun}\approx$\,60\,meV) agrees with theoretically predicted $\delta_{tun}\approx 2|I_{nnn}|\sim$\,70\,meV, if to make use of estimates for $I_{nnn}$ in Ref.\,\onlinecite{Brandow}. Finally, the \emph{d}-\emph{d}, or pair character of this luminescence does explain its strong nonlinear suppression in solid solution Ni$_{0.3}$Zn$_{0.7}$O.  
It is worth noting that self-trapped \emph{d}-\emph{d} excitons can be formed due to a trapping of the oxygen hole borned by the \emph{p}-\emph{d} CT transition on the nearest Ni$^{2+}$ ion. 


\section{Conclusion}
In summary, the luminescence spectra of NiO under VUV and XUV excitation have been investigated. PL spectra show
broad emission bands centered at about 2.3 and 3.2 eV. The PLE spectra recorded in the 4-7\,eV range agree with the RIXS and reflectance data. The PL spectral evolution and lifetime measurements under  VUV excitation reveal that two mechanisms with long and  short decay times, respectively, are responsible for the observed emissions.  These mechanisms are related to $p(\pi )$-$d$ and $d(e_g)$-$d(e_g)$ CT transitions in the range 4-6\,eV, while the most intensive $p(\sigma )$-$d$ CT transition at 7\,ev appears to be a weak if any PL excitation mechanism. The $d(e_g)$-$d(e_g)$ CT transition peaked near 4\,eV and working only in the fast window is the most effective one in the excitation of the high-energy (violet) part of the PL spectrum. For the first time we succeded to distinctly separate contributions of $p(\pi )$-$d$, $p(\sigma )$-$d$, and \emph{d}-\emph{d} CT transitions.   

Making use of the XUV excitation allows us to avoid the predominant role of the surface effects in luminescence and reveal bulk luminescence with puzzling well isolated $I_1$-$I_2$ doublet of very narrow lines with close energies near 3.3\,eV in both NiO and solid solutions Ni$_{x}$Zn$_{1-x}$O. Comparative analysis of the \emph{p}-\emph{d} and \emph{d}-\emph{d} CT transitions,  their relaxation channels,  and luminescence spectra for NiO and Ni$_{0.3}$Zn$_{0.7}$O points to recombination transition in self-trapped  $nnn$ Ni$^{+}$-Ni$^{3+}$ \emph{d}-\emph{d} CT excitons as the only candidate source of unconventional luminescence. To the best of our knowledge it is the first observation of the self-trapping for \emph{d}-\emph{d} CT excitons.




The authors are grateful to R.V. Pisarev, V.I. Anisimov, and A.V. Lukoyanov for discussions and Dr. M. Kirm for help in BW3-experiments. This work was partially supported  by the Russian Federal Agency on Science and Innovation (Grant No. 02.740.11.0217) and RFBR Grant No.~10-02-96032.


\begin{thebibliography}{99}
\bibitem{remark}
Most probably Ksendzov and Drabkin\,\cite{photoconductivity} observed the $p(\pi )$ hole conductivity in a halide-decomposition NiO crystal stimulated by forbidden $p(\pi )$-$d$ ($t_{1g}(\pi)\rightarrow e_g$) CT transition, whereas an attempt to observe photoconductivity in flame-fusion-grown NiO crystals gave negative results up to 7\,eV\,\cite{Powell}, that is up to the energy of allowed $p(\sigma )$-$d$ ($t_{1g}(\sigma)\rightarrow e_g$) CT transition stimulating the $p(\sigma )$ hole conductivity.
\end{thebibliography}
\end{document}